\documentclass[11pt,tightenlines,superscriptaddress,floatfixolumn,english,aps,notitlepage,nofootinbib]{revtex4-1}

\usepackage{amsmath}
\usepackage{esint}
\usepackage{graphicx}
\usepackage{hyperref}
\usepackage{color}
\usepackage{xcolor}
\usepackage[utf8]{inputenc} 
\usepackage{enumitem}
\advance\voffset -0.2in

\begin{document}

\title{
Wounded-quark emission function at the top RHIC energy
}

\author{Micha{\l} Barej}
\email{michal.barej@fis.agh.edu.pl}
\affiliation{AGH University of Science and Technology,
Faculty of Physics and Applied Computer Science,
30-059 Krak\'ow, Poland}

\author{Adam Bzdak}
\email{adam.bzdak@fis.agh.edu.pl}
\affiliation{AGH University of Science and Technology,
Faculty of Physics and Applied Computer Science,
30-059 Krak\'ow, Poland}

\author{Pawe{\l} Gutowski}
\email{pawel.gutowski@fis.agh.edu.pl}
\affiliation{AGH University of Science and Technology,
Faculty of Physics and Applied Computer Science,
30-059 Krak\'ow, Poland}

\begin{abstract}
The wounded nucleon and quark emission functions are extracted for different centralities in d+Au collisions at $\sqrt{s}=200\ \text{GeV}$ using Monte Carlo simulations and experimental data. The shape of the emission function depends on centrality in the wounded nucleon model, whereas it is practically universal (within uncertainties) in the wounded quark model. Predictions for $dN_{ch}/d\eta$ distributions in p+Au and $^3$He+Au collisions are presented.
\end{abstract}

\maketitle

\section{Introduction}
\label{Sec:Introduction}

The idea of a ``wounded'' source is commonly used to model the soft particle production in various hadronic collisions \cite{Bialas:1976ed,Bialas:1977en,Bialas:2012dc}.
In this model, each wounded source populates particles independently of the number of collisions it undergoes.
Typically two models are considered. The wounded nucleon model \cite{Bialas:1976ed} describes a nucleus-nucleus collision as a superposition of nucleon-nucleon interactions. It assumes that each nucleon participating in an inelastic collision is a wounded source. On the other hand, the wounded quark model \cite{Bialas:1977en}, successfully applied to various colliding systems and at various energies \cite{Adler:2013aqf,Adare:2015bua,Bozek:2016kpf,Lacey:2016hqy,Loizides:2016djv,Mitchell:2016jio,Bozek:2017elk,%
Chaturvedi:2016ctn,Zheng:2016nxx}, assumes that a heavy ion collision consists of independent quark-quark collisions and each constituent quark, undergoing inelastic collisions, is a wounded source. The number of wounded sources can assume different values \cite{Loizides:2016djv} and, e.g., in Ref. \cite{abab}, the wounded quark-diquark model was studied.\footnote{We note that wounded quark (diquark) properties can be naturally inferred from a differential proton-proton elastic cross section, as shown in Refs. \cite{abab,Nemes:2012cp,Csorgo:2013bwa,CsorgO:2013kua}.}

In this paper we focus on deuteron-gold (d+Au) collisions at $\sqrt{s}=200\ \text{GeV}$ as experimentally studied by the PHOBOS and PHENIX collaborations at the Relativistic Heavy Ion Collider (RHIC) \cite{Back:2004mr,Aidala:2017pup}. Our goal is to see whether the d+Au data from the PHOBOS collaboration can distinguish between the wounded nucleon and the wounded quark model. Clearly, in both models the average number of particles in d+Au would be comparable since most of the nucleons (in a gold nucleus) collide only once. Thus the only difference in the number of produced particles comes from a deuteron, which typically undergoes several interactions.\footnote{In the wounded nucleon model a nucleon from a deuteron populates particles independently on the number of collisions, whereas in the wounded quark model the number of particles depends on the number of wounded quarks, which clearly depends on the number of collisions.} In order to compare both models, we extracted the wounded source emission functions $F(\eta)$, that is, a pseudorapidity single particle density originating from one wounded source (nucleon, quark, etc.). This object plays a crucial role in hydrodynamic simulations of asymmetric collisions and various studies related to the forward-backward fluctuations and correlations; see, e.g., Refs. \cite{Gazdzicki:2005rr,Adil:2005qn,Bzdak:2009dr,Bozek:2010vz,Bzdak:2012tp,Jia:2014ysa,Pang:2015zrq}. Our goal is to compare $F(\eta)$ in both models and see if and how it changes with centrality.  

We conclude that the shape of the wounded source emission function depends on centrality in the wounded nucleon model; however, it is practically universal for various centralities in the wounded quark model. This suggests that the soft particle production in d+Au collisions is controlled by the number of wounded quarks. A similar problem was studied in Refs. \cite{Bialas:2004su,abab}; however, in these papers the average (over centrality bins) $F(\eta)$ was extracted and the problem of universality of $F(\eta)$ was not investigated.

This paper is organized as follows: In the next section we introduce the wounded nucleon and the wounded quark models, and describe our calculations. Next, we analyze the PHOBOS data and extract the wounded nucleon and quark emission functions. Then, using the derived wounded quark emission function, we predict $dN_{ch}/d\eta$ distributions in p+Au and $^3$He+Au collisions at $\sqrt{s}=200$ GeV. In the last section we present our conclusions.

\section{Two models}
\label{Sec:Wnaq}

As already mentioned in the introduction, we consider two models: the wounded nucleon and the wounded quark models. 
In both models, a single particle pseudorapidity distribution of produced particles can be written as (see Ref. \cite{Bialas:2004su})
\begin{equation} \label{eq:dnch-deta-f}
\frac{dN_{ch}}{d\eta} = w_L F(\eta) + w_R F(-\eta),
\end{equation}
where in the wounded nucleon model $F(\eta)$ is the wounded nucleon emission function, $w_L$ is the average number of the left-going wounded nucleons and $w_R$ is the average number of the right-going wounded nucleons. Both $w_L$ and $w_R$ are calculated at a given centrality class. In the wounded quark model, $F(\eta)$ is the wounded quark emission function, and $w_L$ and $w_R$ are the average numbers of the left- and right-going wounded quarks, respectively. If $w_L \neq w_R$,\footnote{We note that if $w_L = w_R$ we can extract only $F(\eta ) + F(-\eta)$.} the wounded source emission function can be extracted for each centrality and is given by
\begin{equation} \label{eq:fragm-fun}
F(\eta) = \frac{1}{2} \left[ \frac{N(\eta) + N(-\eta)}{w_L + w_R} + \frac{N(\eta) - N(-\eta)}{w_L - w_R} \right],
\end{equation}
where $N(\eta) := dN_{ch}/d\eta$ and is taken from the PHOBOS measurement on d+Au collisions at $\sqrt{s} = 200\ \text{GeV}$ \cite{Back:2004mr}, which covers the center of mass pseudorapidity range $|\eta| \le 5.3$. Note that Eq. (\ref{eq:fragm-fun}) is not applicable for symmetric (or for very peripheral) collisions because the second part of this equation becomes indefinite in such cases. 

\subsection{Wounded nucleon model}

In the wounded nucleon model, each wounded nucleon\footnote{By definition, a wounded nucleon undergoes at least one inelastic collision.} populates soft particles independently of the number of collisions it undergoes. 

In our Monte Carlo calculation, the impact parameter squared $b^2$ is drawn from a uniform distribution in an interval of $[0, b_{max}^2]$ with $b_{max} = 15\ \text{fm}$. The positions of nucleons in the gold nucleus are drawn according to the Woods-Saxon distribution \cite{DeJager:1987qc,Loizides:2014vua}
\begin{equation} \label{eq:woods-saxon}
\varrho(\vec{r}) = \varrho_0 \frac{1}{1 + \exp \left( \frac{r - R}{a} \right)},
\end{equation}
where $r = |\vec{r}|$ is the distance from the nucleus center, $\varrho_0$ is the nucleon density, $R = 6.38\ \text{fm}$ is the nuclear radius and $a = 0.535\ \text{fm}$ is the skin depth.\footnote{We checked that introducing a minimal distance between nucleons $d_{min} = 0.4\ \text{fm}$ has a negligible effect on our results.} For a deuteron, the proton's position is described by the Hulthen form
\begin{equation} \label{eq:hulthen}
\varrho(\vec{r}) = \varrho_0  \left( \frac{e^{-Ar} - e^{-Br}}{r} \right)^2,
\end{equation}
where $A = 0.457\ \mathrm{fm}^{-1}$,  $B = 2.35\ \mathrm{fm}^{-1}$, and the neutron is placed opposite to the proton \cite{hulthen,Loizides:2014vua}. 

Next, for each nucleon from the left-going nucleus it is checked whether it collides with each nucleon from the right-going nucleus using a probability function. The simple Heaviside step function was used, namely, two nucleons (one from each nucleus) collide if a transverse distance, $d$, between them is $d \le \sqrt{ \sigma_{nn} / \pi }$.\footnote{We checked that the collision probability function given by a normal distribution results in a very similar wounded quark emission function.} The inelastic nucleon-nucleon cross section was taken to be $\sigma_{nn} = 41\ \mathrm{mb}$ ($\sqrt{s}=200\ \text{GeV}$).

To define centrality through the number of produced particles, each wounded nucleon emits charged particles (independently on the number of collisions it underwent). In our approach, particles are emitted according to a negative binomial distribution with the mean number of particles $\langle n \rangle = 5$ and $k = 1$ \cite{Ansorge:1988kn}, where $k$ measures the deviation from Poisson distribution.\footnote{There are $w = w_L + w_R$ wounded nucleons and the superposition of $w$ independent negative binomial distributions with same $\langle n \rangle$ and $k$ is also a negative binomial distribution with parameters $w \langle n \rangle$ and $w k$.}

The simulation results have been divided into 0-20\%, 20-40\%, 40-60\%, 60-80\%, and 80-100\% centrality bins. For each centrality class, the number of wounded nucleons in the left-going nucleus, $w_L$, and the number of wounded nucleons in the right-going nucleus, $w_R$, were calculated to complete Eq. (\ref{eq:fragm-fun}).

\subsection{Wounded quark model}

In the wounded quark model, particle production is controlled by quark-quark collisions rather than nucleon-nucleon collisions. In order to use Eq. (\ref{eq:fragm-fun}), it was necessary to find $w_L$ and $w_R$, the numbers of wounded quarks from the left and the right-going nucleus, respectively. 

The positions of three constituent quarks around the center of each nucleon are drawn according to 
\begin{equation} \label{eq:quark-density}
\varrho(\vec{r}) = \varrho_0 \exp \left(- \frac{r}{a} \right),
\end{equation}
where $a = \frac{r_p}{\sqrt{12}}$ with $r_p = 0.81 \ \text{fm}$ being the proton's radius \cite{Hofstadter:1956qs,Adler:2013aqf}.\footnote{Three quarks are shifted so that their center of mass is located in the center of a nucleon. After this procedure, the quarks are no longer consistent with $\varrho(\vec{r})$. To deal with this problem, we changed $\varrho(\vec{r})$ into $\widetilde{\varrho}(\vec{r}) = \varrho_0 \exp \left(- Cr/a \right)$. $C$ was determined in simulation by the trial and error method and we obtained $C=0.82$.} The locations of nucleons are drawn according to Eqs. (\ref{eq:woods-saxon}) and (\ref{eq:hulthen}).

The calculations in the wounded quark model are carried out in a way analogous to that of the wounded nucleon model. For each quark from the left-going nucleus, it is checked whether it collides with each quark from the right-going nucleus according to the Heaviside step function with $d \le \sqrt{ \sigma_{qq}/ \pi }$, where $\sigma_{qq}$ is the inelastic quark-quark collision cross section. We determined $\sigma_{qq}$ using the trial and error method in simulation. We were looking for the value $\sigma_{qq}$, for which $\sigma_{nn} = \int_{0}^{2 \pi} d\varphi \int_{0}^{+ \infty} db P(b) b$, where $P(b)$ is a probability of proton-proton collision with the impact parameter $b$, is equal to the desired value of 41 mb. We obtained $\sigma_{qq} \simeq 7 \ \text{mb}$.

It was assumed that each wounded quark emits charged particles with respect to negative binomial distribution with $k_q$ and $\langle n_q \rangle$ parameters. Taking p+p collisions into consideration, one observes that at $\sigma_{nn} = 41\ \text{mb}$ the average number of wounded quarks is about 1.3 per one wounded nucleon (this value depends on $\sqrt{s}$). Therefore, we take $k_q = k_p/1.3$, $\langle n_q \rangle = \langle n_p \rangle / 1.3$, where $k_p = 1$ and $\langle n_p \rangle = 5$ are the parameters of NBD for protons used in the wounded nucleon model calculations.

Finally, the numbers of wounded quarks, $w_L$ and $w_R$, are calculated for each centrality class in the same way as before.

\section{Results}
\label{Sec:Results}

In this section we extract the wounded nucleon and quark emission functions $F(\eta)$, using Eq. (\ref{eq:fragm-fun}). The pseudorapidity distribution of charged particles $dN_{ch}/d\eta$ is taken from the PHOBOS measurement \cite{Back:2004mr}. 

\subsection{Wounded nucleon emission function}

The mean numbers of wounded nucleons in d+Au collisions at $\sqrt{s}=200\ \text{GeV}$ obtained in our Monte Carlo simulation and used for further calculations are presented in Tab. \ref{table:d-au-nn}. 

\begin{table}[h!]
\begin{tabular}{|r|r|r|r|r|r|} \hline
 & min-bias & 0-20\% & 20-40\% & 40-60\% & 60-80\% \\ \hline
d & 1.61 & 1.96 & 1.85 & 1.65 & 1.38 \\ \hline
Au & 6.69 & 13.65 & 8.96 & 5.63 & 3.04 \\ \hline
\end{tabular}
\caption{The mean number of wounded nucleons for different centrality classes in d+Au collisions at $\sqrt{s} = 200\  \text{GeV}$.}\label{table:d-au-nn}
\end{table}

\begin{figure}[t]
\begin{center}
\includegraphics[scale=0.45]{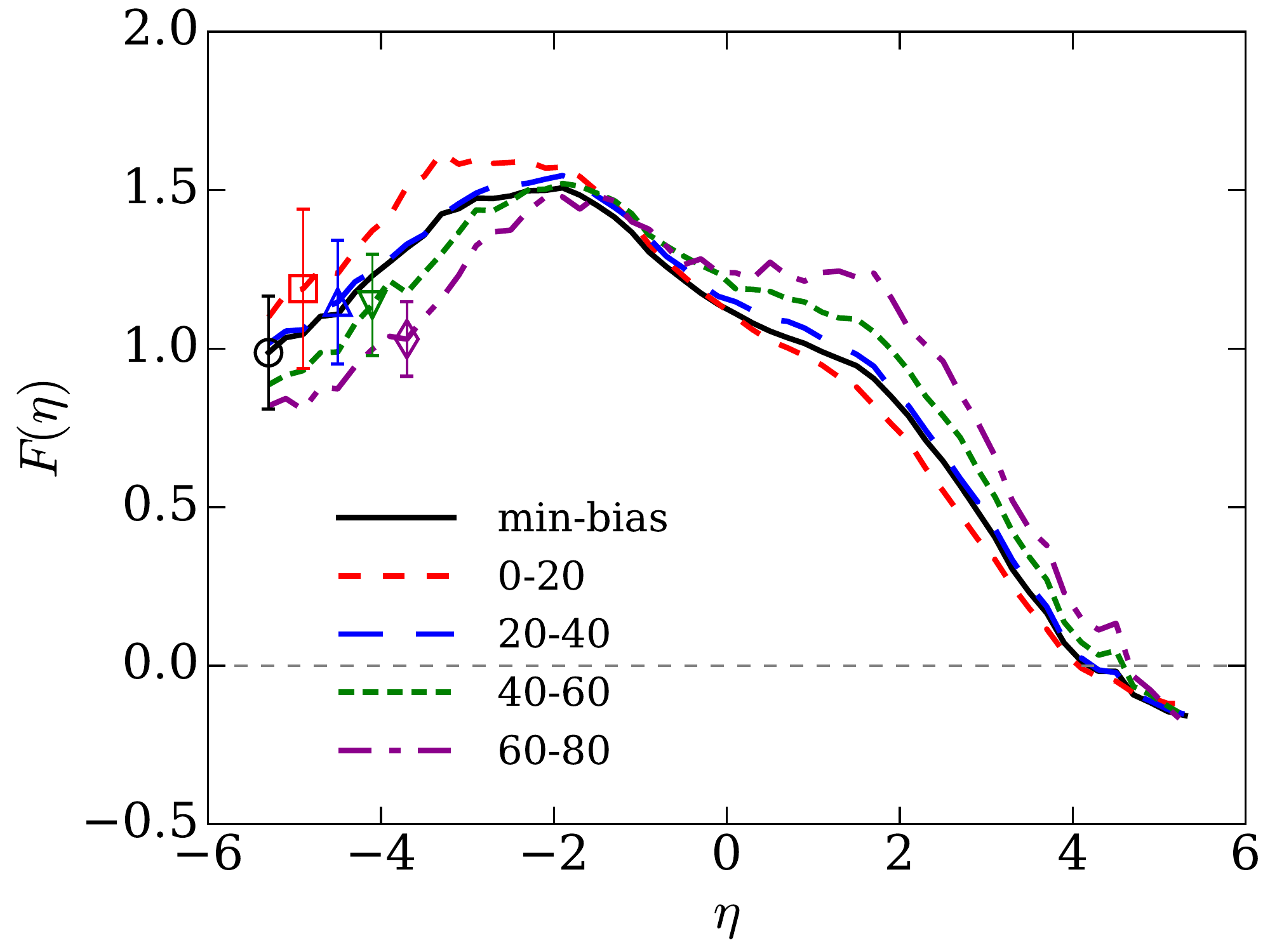} 
\end{center}
\par
\vspace{-5mm}
\caption{The wounded nucleon emission functions in different centrality classes extracted from the wounded nucleon model using $dN_{ch}/d\eta$ from \cite{Back:2004mr} and the numbers of wounded nucleons from our Monte Carlo simulation of d+Au collisions at $\sqrt{s} = 200\ \text{GeV}$. Each wounded nucleon emits particles according to a negative binomial distribution with $k = 1$ and $\langle n \rangle = 5$.}
\label{fig:nn}
\end{figure}

Using the values from Tab. \ref{table:d-au-nn}, the wounded nucleon emission function was extracted according to Eq. (\ref{eq:fragm-fun}). This is presented in Fig. \ref{fig:nn}. Each line represents a different centrality bin. The uncertainty of the emission function, $F(\eta)$, was calculated using the uncertainties of $N(\eta)$ and $N(-\eta)$.\footnote{The errors represent the systematic uncertainties of $N(\eta)$ 
and they are not expected to influence the shape of $F(\eta)$ but its overall normalization only.} For clarity, we show errors in the limited range of $\eta$. We note that, using the numbers of wounded nucleons estimated by PHOBOS \cite{Back:2004mr}, we obtained virtually identical wounded nucleon emission functions.

As seen in Fig. \ref{fig:nn}, the shape of the wounded nucleon emission function differs with centrality. The negative value of $F(\eta)$ (for $\eta > 4$) has obviously no physical sense indicating that our model is not applicable for $\eta > 4$. This is not surprising since large pseudorapidity values are influenced by the fragmentation physics, which is not included in our model.

\subsection{Wounded quark emission function}

\begin{figure}[t]
\begin{center}
\includegraphics[scale=0.45]{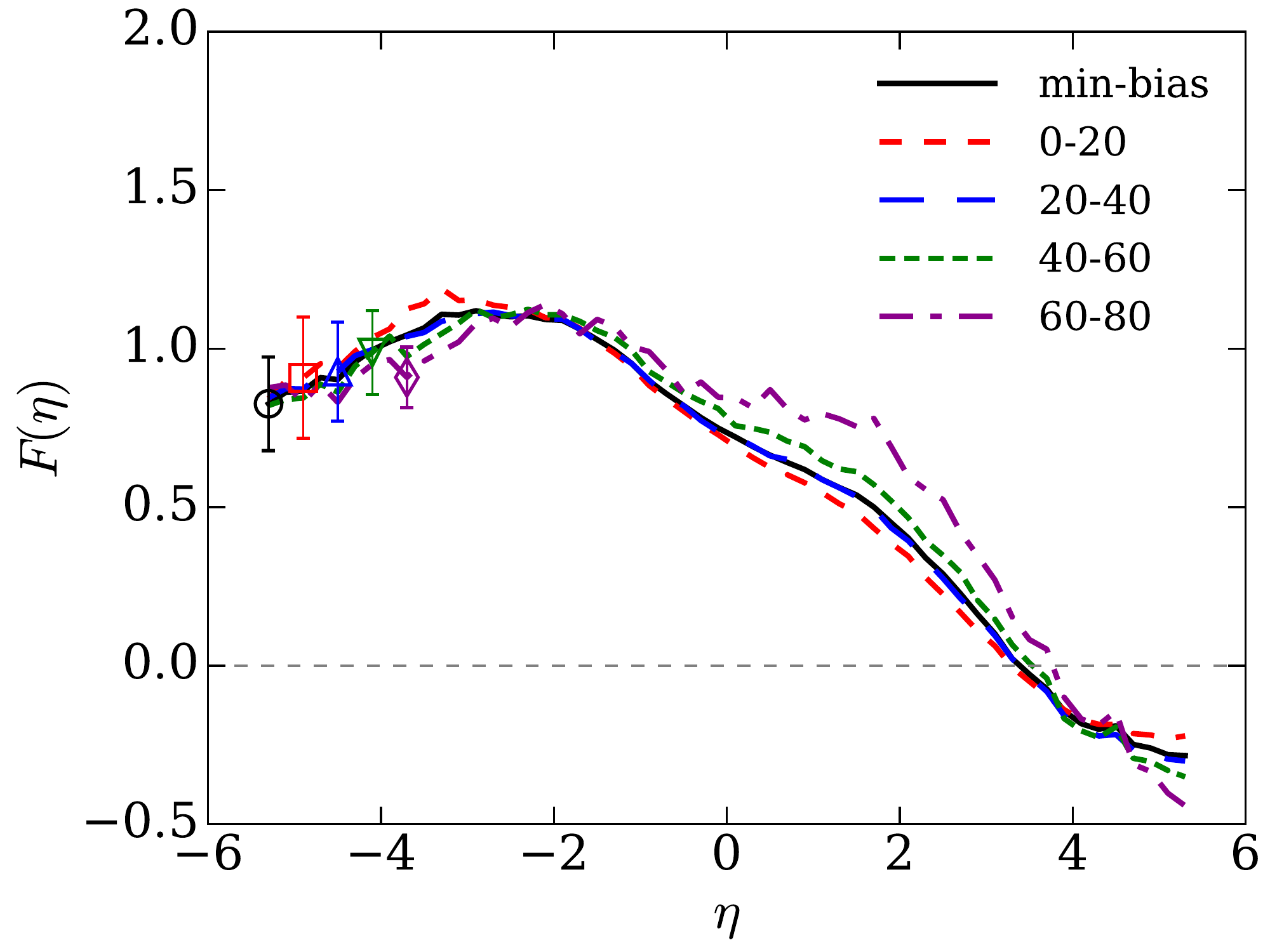} 
\end{center}
\par
\vspace{-5mm}
\caption{The wounded quark emission functions in different centrality bins extracted from the wounded quark model using $dN_{ch}/d\eta$ from \cite{Back:2004mr} and the numbers of wounded quarks from our Monte Carlo simulation of d+Au collisions at $\sqrt{s} = 200\ \text{GeV}$. Each wounded quark emits particles according to a negative binomial distribution with $k_q = 1/1.3$ and $\langle n_q \rangle = 5/1.3$.}
\label{fig:qq-1-5}
\end{figure}

Our next step is to extract the wounded quark emission function. The calculated mean numbers of wounded quarks are presented in Tab. \ref{table:d-au-qq}, whereas in Fig. \ref{fig:qq-1-5} we show the extracted emission functions. 

\begin{table}[h!]
\begin{tabular}{|r|r|r|r|r|r|} \hline
 & min-bias & 0-20\% & 20-40\% & 40-60\% & 60-80\% \\ \hline
d & 3.73 & 5.63 & 4.93 & 3.86 & 2.61 \\ \hline
Au & 8.97 & 19.01 & 12.19 & 7.41 & 3.87 \\ \hline
\end{tabular}
\caption{The mean numbers of wounded quarks for different centrality bins in d+Au collisions at $\sqrt{s} = 200\  \text{GeV}$.}\label{table:d-au-qq}
\end{table}

As shown in Fig. \ref{fig:qq-1-5}, the shape of $F(\eta)$ is rather similar for different centrality classes. In other words, to understand the d+Au data on $dN_{ch}/d\eta$ we need one emission function and the main difference between different centralities comes from different values of $w_L$ and $w_R$. This suggests that the wounded quarks are indeed more suited to describe soft particle production in d+Au collisions.
We also note that the wounded quark emission function is physically meaningful for $|\eta| \le 3$ because, for large 
$|\eta|$ (fragmentation regions), contributions from unwounded quarks (within wounded nucleons) become significant \cite{abab}.

The universal character of the wounded quark emission function is not unexpected.
As already emphasized in the introduction, the wounded quark model describes rather well 
the mid-rapidity multiplicities in A+A collisions for all centralities and across 
a broad range of energies.

\begin{figure}[t]
\begin{center}
\includegraphics[scale=0.45]{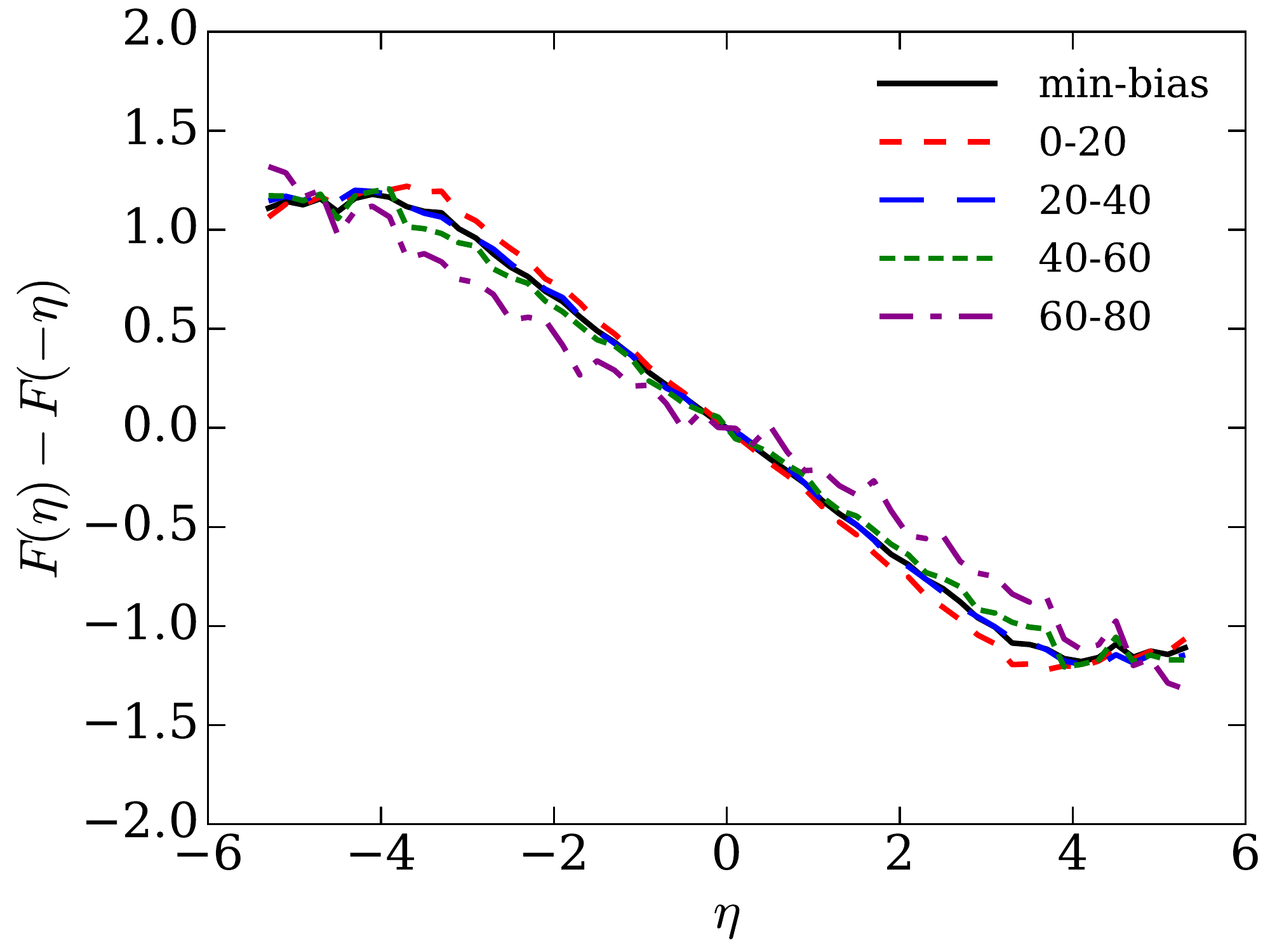} 
\end{center}
\par
\vspace{-5mm}
\caption{The antisymmetrized wounded quark emission functions, $F(\eta)-F(-\eta)$, for different centrality bins, as extracted from d+Au collisions at $\sqrt{s} = 200\ \text{GeV}$.}
\label{fig:qq-diff}
\end{figure}

To further test the shape of the wounded quark emission functions,  
we plot in Fig. \ref{fig:qq-diff} the difference $F(\eta)-F(-\eta)$ for different centrality classes. We observe that, indeed, the shape is practically independent on centrality and, apart from the fragmentation regions, $F(\eta)-F(-\eta) \sim \eta$.

\subsection{Predictions for p+Au and $^3$He+Au}

The results presented in Fig. \ref{fig:qq-1-5} encouraged us to make predictions for p+Au and $^3$He+Au collisions. The mean numbers of wounded quarks in p+Au collisions are presented in Tab. \ref{table:p-au-qq-nn}.

\begin{table}[h!]
\begin{tabular}{|r|r|r|r|r|r|} \hline
 & min-bias & 0-20\% & 20-40\% & 40-60\% & 60-80\% \\ \hline
p & 2.33 & 2.93 & 2.78 & 2.48 & 1.97 \\ \hline
Au & 5.84 & 11.32 & 7.51 & 5.09 & 3.04 \\ \hline
\end{tabular}
\caption{The mean numbers of wounded quarks for various centrality bins in p+Au collisions at $\sqrt{s} = 200\  \text{GeV}$.}\label{table:p-au-qq-nn}
\end{table}

In the case of $^3$He+Au collisions, the positions of nucleons in $^3$He nuclei have been taken from \cite{Carlson:1997qn}. The locations of quarks relative to each nucleon have been drawn according to Eq. (\ref{eq:quark-density}) (see also a corresponding footnote). The mean numbers of wounded quarks have been determined for each centrality bin and are presented in Tab. \ref{table:he3-au-qq-nn}.

\begin{table}[h!]
\begin{tabular}{|r|r|r|r|r|r|} \hline
 & min-bias & 0-20\% & 20-40\% & 40-60\% & 60-80\% \\ \hline
$^3$He & 5.39 & 8.52 & 7.57 & 5.68 & 3.34 \\ \hline
Au & 11.82 & 26.15 & 16.65 & 9.51 & 4.36 \\ \hline
\end{tabular}
\caption{The mean numbers of wounded quarks for various centrality bins in $^3$He+Au collisions at $\sqrt{s} = 200\  \text{GeV}$.}\label{table:he3-au-qq-nn}
\end{table}

\begin{figure}[t]
\begin{center}
\includegraphics[scale=0.45]{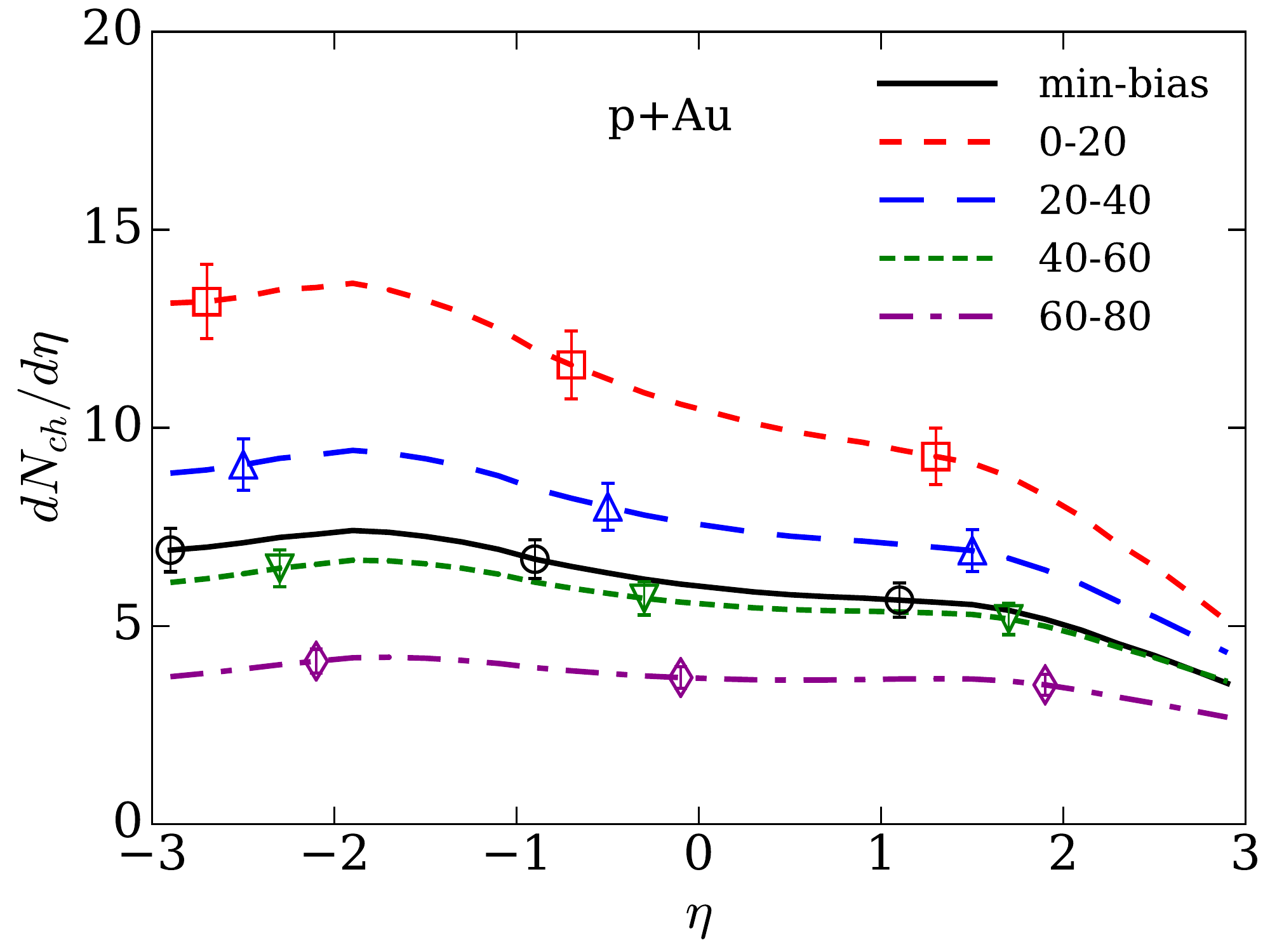} 
\end{center}
\par
\vspace{-5mm}
\caption{$dN_{ch}/d\eta$ predicted for p+Au collisions at $\sqrt{s}$=200 GeV in the wounded quark model.} 
\label{fig:p-au-no}
\end{figure}

\begin{figure}[t]
\begin{center}
\includegraphics[scale=0.45]{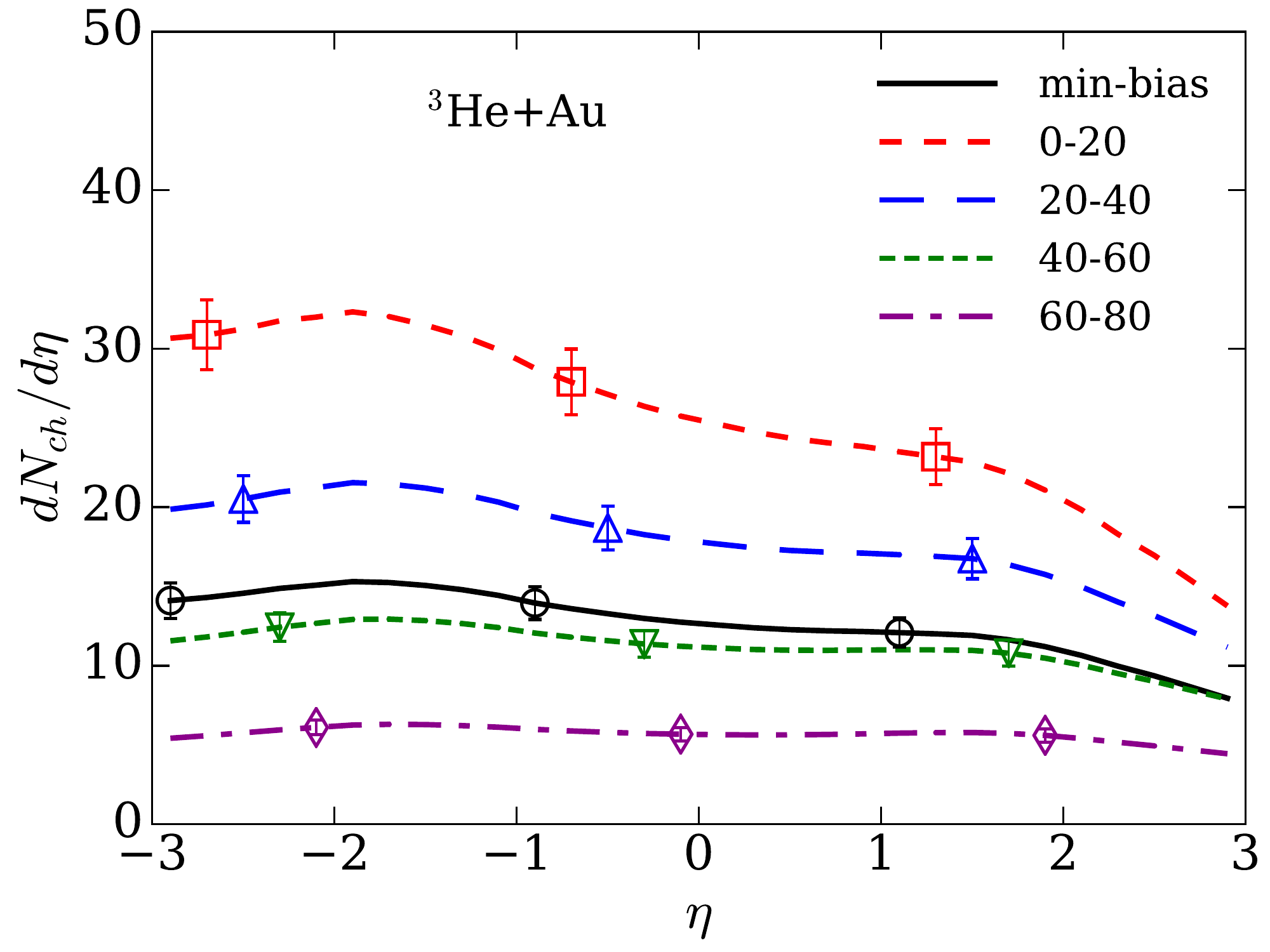} 
\end{center}
\par
\vspace{-5mm}
\caption{$dN_{ch}/d\eta$ predicted for $^3$He+Au collisions at $\sqrt{s}$=200 GeV in the wounded quark model.} 
\label{fig:he3-au-no}
\end{figure}

Assuming that the wounded quark emission function $F(\eta)$ is universal not only for different centrality bins but also for various colliding nuclei, using Eq. (\ref{eq:dnch-deta-f}), we predict $dN_{ch}/d\eta$ distributions in p+Au and $^3$He+Au collisions. The results for p+Au as well as for $^3$He+Au cut to the region $|\eta| \le 3$ are shown in Figs. \ref{fig:p-au-no} and \ref{fig:he3-au-no}, respectively. Here the minimum-bias wounded quark emission function was used; see Fig. \ref{fig:qq-1-5}. The uncertainties of $dN_{ch}/d\eta$ were calculated using the uncertainties of extracted $F(\eta)$.


\section{Conclusions}
\label{Sec:Conclusion}

Our conclusions can be formulated as follows:

\begin{enumerate}[label=(\roman*)]
  \item Two models, the wounded nucleon and the wounded quark models, have been applied to simulate d+Au collisions at $\sqrt{s}=200\ \text{GeV}$ and to describe the process of soft particle production. The pseudorapidity distributions of charged particles $dN_{ch}/d\eta$ have been taken from PHOBOS measurement \cite{Back:2004mr}. 
  
  \item The wounded nucleon and quark emission functions $F(\eta)$ have been extracted to compare both models. In the wounded nucleon model, the shape of $F(\eta)$ differs for various centrality bins. In contrast, in the wounded quark model, the extracted functions are practically universal for all centrality classes at $|\eta| \le 3$. This observation suggests that the d+Au collision is better described by the wounded quark model and particle production takes place at the quark level.
  
  \item There are many different models used to describe p+A and A+A collisions; see, e.g., LEXUS \cite{Jeon:1997bp}, which is a simple extrapolation of nucleon-nucleon to nucleus-nucleus collisions and conceptually is not far from our framework. See also models like HIJING \cite{Wang:1991hta}, UrQMD \cite{Bleicher:1999xi}, AMPT \cite{Lin:2004en}, or EPOS \cite{Pierog:2009zt}, which are rather advanced parton-based Monte Carlo tools. Our approach is quite different. Instead of trying to fit the d+Au data with a certain number of parameters, we extracted the wounded nucleon and quark emission functions in a parameter-free way.
  
  \item Assuming the extracted quark emission function can be applied to various collision types at the same $\sqrt{s}$, the distributions $dN_{ch}/d\eta$ have been predicted for p+Au and $^3$He+Au at $\sqrt{s}=200\ \text{GeV}$. Hopefully, they can be verified experimentally.
  
  \item In this work we extracted the single particle wounded quark emission function. It would be desired to learn how $F(\eta)$ fluctuates from event to event. The recent measurement of $\langle a_1^{2} \rangle$ \cite{Bzdak:2012tp} by the ATLAS collaboration may shed some light on this problem \cite{Aaboud:2016jnr}.
  
  \item For future research, it would be interesting to verify the model at the LHC energies. However, at this moment such an exercise cannot be done since the available data on p+Pb collisions are strongly dependent on the centrality definition. Also, it would be desired to verify the model at various energies in d+Au interactions as currently studied by the PHENIX collaboration.
  
  \item Finally, it would be interesting to interpret $F(\eta)$ in the color glass framework \cite{Iancu:2002xk,Gelis:2010nm}, where a longitudinal structure of the color flux tubes may be not far from what is presented in Fig. \ref{fig:qq-1-5}. This and other related questions are currently under our investigation. 
\end{enumerate}

 \begin{acknowledgements}
We thank A. Bia{\l}as for helpful comments. We are grateful to A. Szkudlarek for useful discussions and his contribution at the early stage of this work. 
A.B. is partially supported by the Faculty of Physics and Applied Computer Science AGH UST statutory tasks within subsidy of Ministry of Science and Higher Education, and by the National Science Centre, Grant No. DEC-2014/15/B/ST2/00175.
 \end{acknowledgements}

\appendix
\section{PHENIX request}

\begin{figure}[t]
\begin{center}
\includegraphics[scale=0.4]{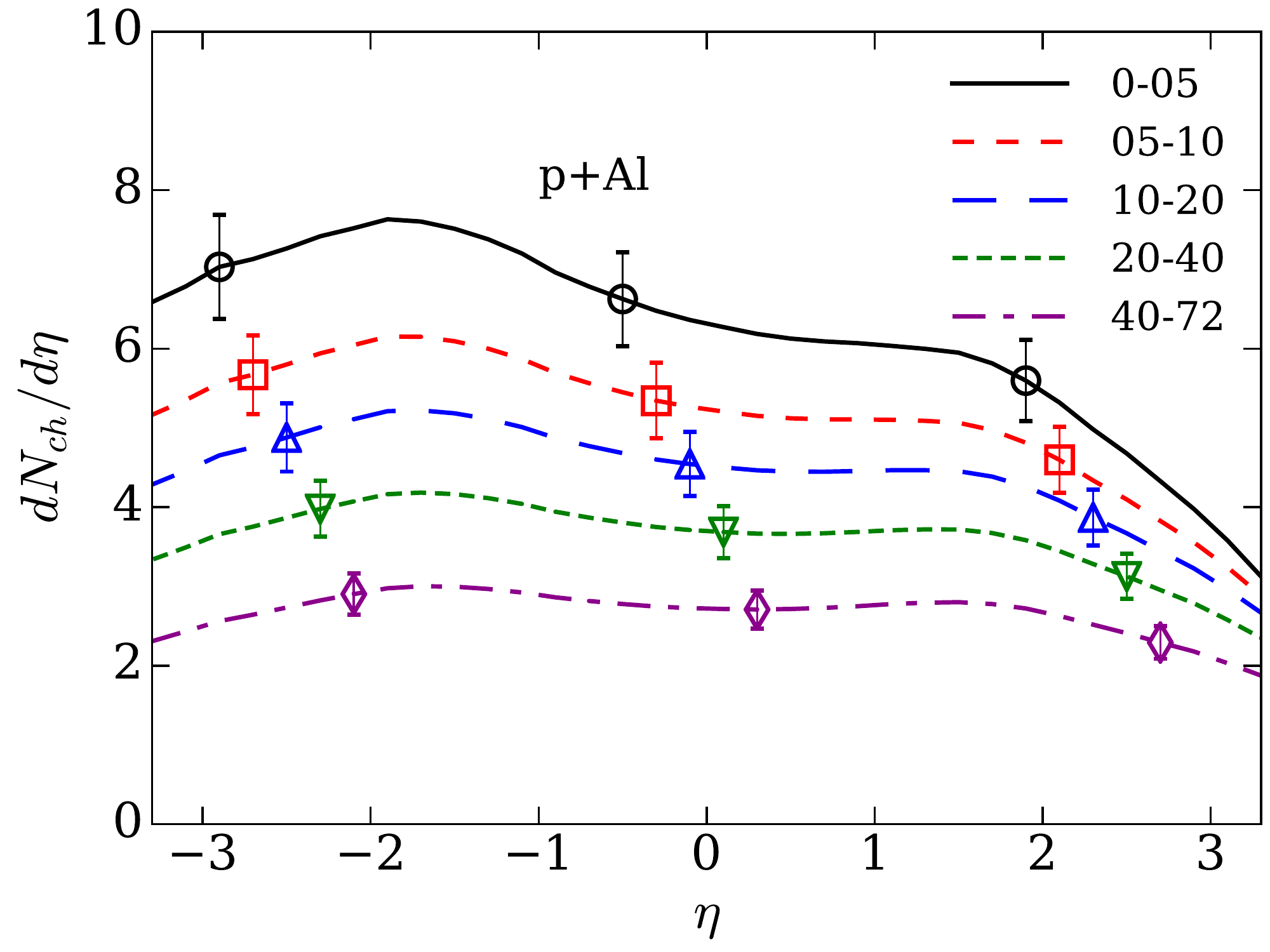} 
\includegraphics[scale=0.4]{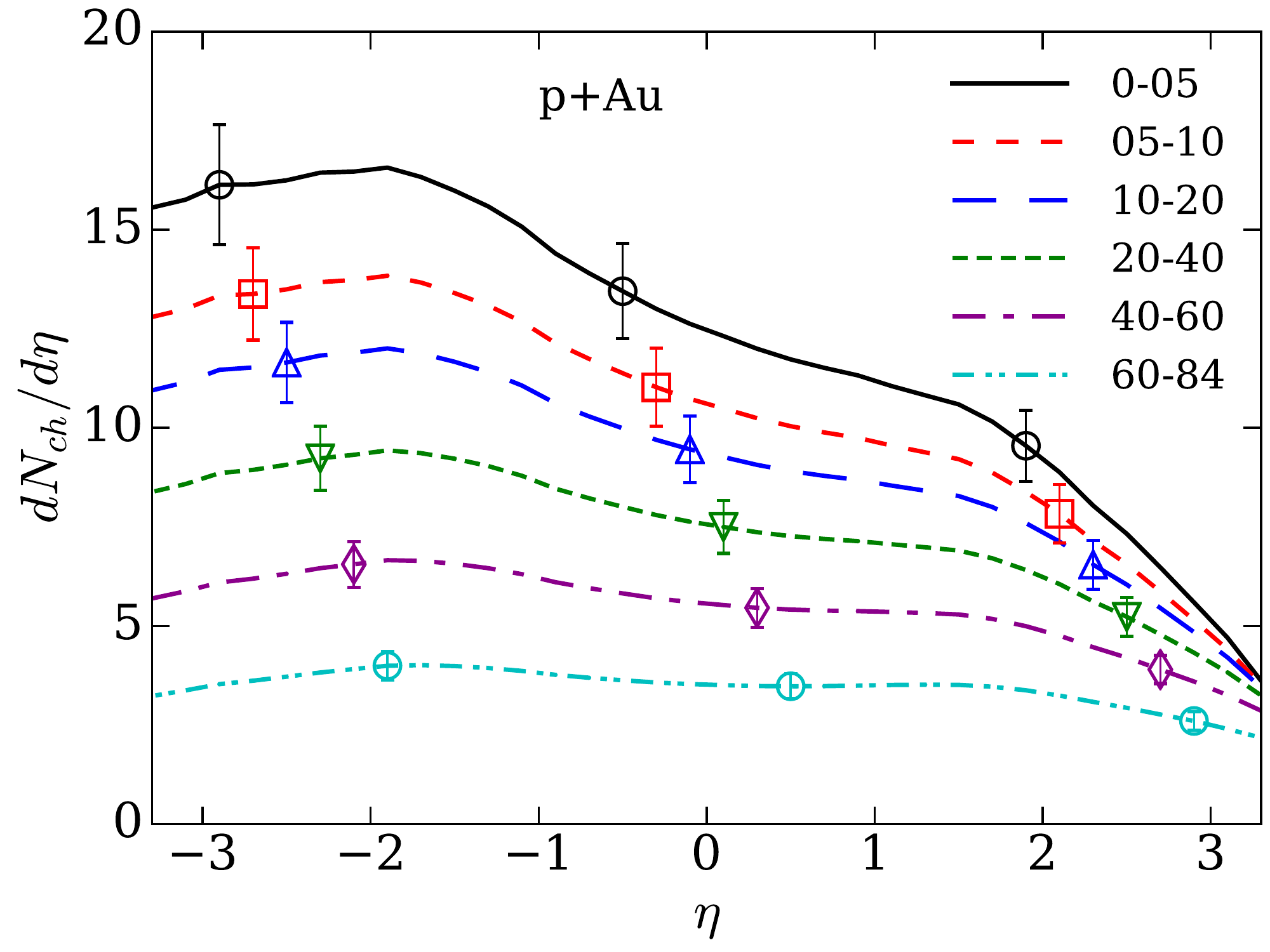} 
\includegraphics[scale=0.4]{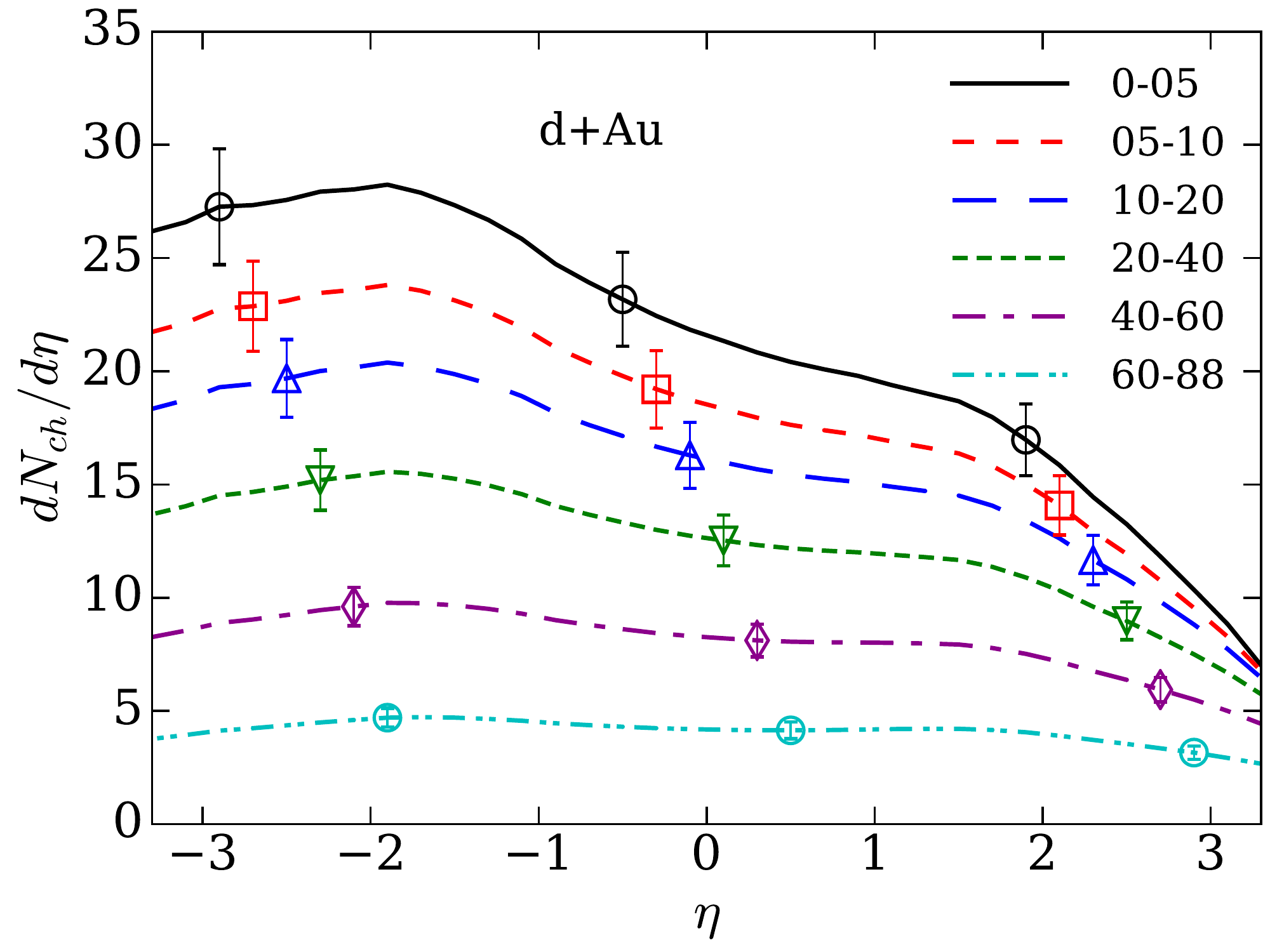} 
\includegraphics[scale=0.4]{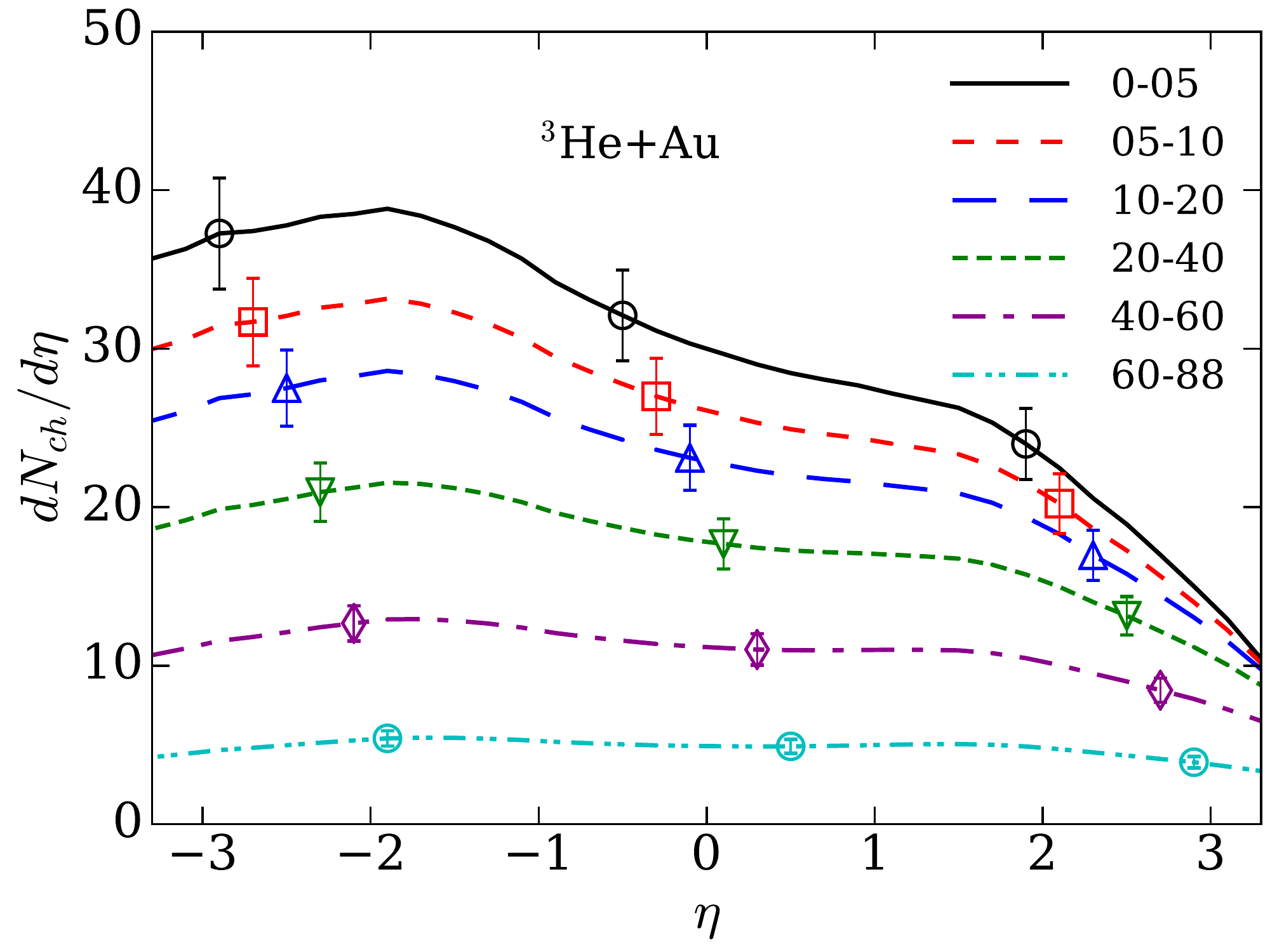} 
\end{center}
\par
\vspace{-5mm}
\caption{$dN_{ch}/d\eta$ predicted for p+Al, p+Au, d+Au, and $^3$He+Au collisions at $\sqrt{s}$=200 GeV in the wounded quark model. The errors are dominated by the PHOBOS errors plus we added $5\%$ (in quadratic) related to the number of wounded quarks.}
\label{fig:phenix-all}
\end{figure}

In Fig. \ref{fig:phenix-all} we present the predicted $dN_{ch}/d\eta$ distributions for p+Al, p+Au, d+Au, and $^3$He+Au at various centralities as requested by the PHENIX Collaboration. These plots are not included in the published manuscript.

\end{document}